\documentclass[12pt,notoc]{JHEP3}

\usepackage{amsmath,amssymb}

\DeclareMathOperator{\Hilb}{Hilb}
\DeclareMathOperator{\Tr}{Tr}

\title{Abelian Vortices on Nodal and Cuspidal Curves}

\author{Toshiya Kawai\\
Research Institute for Mathematical Sciences,\\
 Kyoto University, Kyoto,  Kyoto 606--8502, Japan}

\date{}

\abstract{
We compute the Euler characteristics of the  moduli spaces of abelian vortices on   curves with nodal and cuspidal singularities. This generalizes  our  previous work  where only nodes were taken into account.  The result we obtain is again consistent with the expected  reconciliation between  the vortex picture of $D2$-$D0$ branes and the proposal by  Gopakumar and Vafa.
}

\begin{document}

\newcommand{\CC}{{\mathbb{C}}}
\newcommand{\GG}{{\mathbb{G}}}
\newcommand{\RR}{\mathbb{R}}
\newcommand{\QQ}{\mathbb{Q}}
\newcommand{\ZZ}{\mathbb{Z}}
\newcommand{\HH}{\mathbb{H}}
\newcommand{\NN}{\mathbb{N}}
\newcommand{\PP}{\mathbb{P}}
\newcommand{\abs}[1]{\lvert#1\rvert}
\newcommand{\inv}[1]{\frac{1}{#1}}

%%%%%%%%%%%%%%%%%%%%%%%%%%%%%%%%%%%%%%%%%%%%%%%%
\section{Introduction}
Let $C$ be a nonsingular complex projective curve of genus $g$.
The moduli space of abelian vortices on  $C$  is well-known to be
described by   the $d$-fold symmetric product $C^{(d)}$ where $d$ is the amount of magnetic flux. Its Euler characteristic $\chi(C^{(d)})$ can be 
computed  via  the generating function \cite{Macdonald:1962ya}
\begin{equation}\label{symproduct}
\sum_{d=0}^\infty\chi(C^{(d)})y^{d+1-g}=(y^{\inv{2}}-y^{-\inv{2}})^{2g-2}
\end{equation}
where $0<\abs{y}<1$ is assumed.
 In view of the existence of the Abel-Jacobi map from $C^{(d)}$ to the Jacobian $J(C)$  it is not unreasonable to expect  a close relation
  between the two. The total complex  cohomology ring $H^*(J(C))$ is an   $\mathfrak{sl}_2$ module under 
  the Lefschetz  $\mathfrak{sl}_2$ action. If we denote
the Cartan generator of the  $\mathfrak{sl}_2$ by $\mathsf{H}$, we have
\begin{equation}\label{jacobian}
\Tr_{H^*(J(C))}(-1)^{\mathsf{H}}y^{\mathsf{H}}=(-1)^g(y^{\inv{2}}-y^{-\inv{2}})^{2g}.
\end{equation}
Then, we observe that \eqref{symproduct} and \eqref{jacobian} coincide up to  a simple factor $(-1)^g(y^{\inv{2}}-y^{-\inv{2}})^{2}$.

Abelian vortices on a curve are expected to describe  the bound system of a single $D2$-brane  coupled to $D0$-branes.  In this context, the above   observation, though it may look accidental, is crucial for the  reconciliation  between the vortex picture of $D2$-$D0$ branes \cite{Kawai:2000pr}
and the proposal  by  Gopakumar and Vafa \cite{Gopakumar:1998qr}. Recall that the latter   is  motivated by   an effective theory consideration\footnote{In the sense of string compactification.}  and tries to interpret the Lefschetz  $\mathfrak{sl}_2$ action on the cohomologies of the Jacobian   as the {\em half\/} of space-time Lorentz symmetry.   
In many interesting and important cases though, the curve around which the $D2$-brane is wrapping can be singular and a priori one is not sure  if the same  kind of simple relation holds. 
Nevertheless, such a relation seems to be  required  if one believes in  the compatibility of the two pictures.
In  \cite{Kawai:2004fh},  we studied this issue when the singularities of the curve are nodes and found that the two expressions are again simply related  as in the nonsingular case.
In this short note,  we modestly extend this result by  additionally allowing  cusps on the curve.
 See  \cite{Kawai:2004fh} for more on the motivation behind the present work and the background materials. 
 
 The main computation for abelian vortices on nodal and cuspidal curves is given in  \S 2.  We compare this result with  the Gopakumar-Vafa type expression for the compactified Jacobians  in \S 3.

\section{Abelian vortices on nodal and cuspidal curves}

Let $C$ be an integral  complex  projective curve of arithmetic genus $g$ having $a$ nodes and $b$ cusps as its only singularities. 
We denote by   $C^{[d]}$   the  Hilbert scheme of zero-dimensional subschemes of length $d$  on $C$.  One may regard  $C^{[d]}$ as the moduli space of vortices on $C$.
Then our   claim is  that
\begin{equation}
\label{formula}
\begin{split}
&\sum_{d=0}^\infty \chi(C^{[d]})y^{d+1-g}\\
&\quad =(y^{\inv{2}}-y^{-\inv{2}})^{2g-2}\left(1+\inv{(y^{\inv{2}}-y^{-\inv{2}})^{2}}\right)^a
\left(1+\frac{2}{(y^{\inv{2}}-y^{-\inv{2}})^{2}} \right)^b
\end{split}
\end{equation}
for $0<\abs{y}<1$.

In order to prove this,  we first gather  relevant materials on local punctual Hilbert schemes  at singularities.
The local punctual Hilbert scheme $\Hilb_{ \propto}^\ell$ at a node parametrizes ideals of colength $\ell$ in
$\CC[\![ \mathtt{x},\mathtt{y}]\!]/( \mathtt{x}\mathtt{y})$.  If $\ell>1$, such ideals are given by  \cite{Ran:2005iw}
\begin{equation}
\begin{split}
I_i^\ell(u_i)&=(\mathtt{y}^i+u_i \mathtt{x}^{\ell-i}),\quad (u_i\in \CC^\times, i=1,\dots,\ell-1),\\
Q_i^\ell &=(\mathtt{x}^{\ell-i+1},\mathtt{y}^i), \quad (i=1,\dots,\ell)
\end{split}
\end{equation}
with the relations $\lim_{u_i\to 0} I_i^\ell(u_i)=Q_i^\ell $  and $\lim_{u_i\to \infty} I_i^\ell(u_i)=Q_{i+1}^\ell $.
Hence
$\Hilb_{ \propto}^\ell$ with $\ell>1$ is a chain of $\ell-1$ rational curves configured as  \cite{Ran:2005iw}:
\medskip
\ifx\JPicScale\undefined\def\JPicScale{0.8}\fi
\unitlength \JPicScale mm
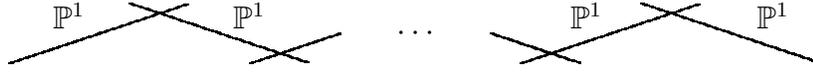
\begin{center}
\begin{picture}(135,10)(0,0)
\linethickness{0.2mm}
\multiput(0,0)(0.36,0.12){83}{\line(1,0){0.36}}
\linethickness{0.2mm}
\multiput(20,10)(0.36,-0.12){83}{\line(1,0){0.36}}
\linethickness{0.2mm}
\multiput(40,0)(0.36,0.12){42}{\line(1,0){0.36}}
\linethickness{0.2mm}
\multiput(80,5)(0.36,-0.12){42}{\line(1,0){0.36}}
\linethickness{0.2mm}
\multiput(85,0)(0.36,0.12){83}{\line(1,0){0.36}}
\linethickness{0.2mm}
\multiput(105,10)(0.36,-0.12){83}{\line(1,0){0.36}}

\put(10,8){\makebox(0,0)[cc]{$\PP^1$}}

\put(40,8){\makebox(0,0)[cc]{$\PP^1$}}

\put(127,8){\makebox(0,0)[cc]{$\PP^1$}}

\put(96,8){\makebox(0,0)[cc]{$\PP^1$}}

\put(68,5){\makebox(0,0)[cc]{$\cdots$}}

\end{picture}
\end{center}
\medskip
The only colength one  ideal is $Q_1^1=(\mathtt{x},\mathtt{y})$. Hence $\Hilb_{ \propto}^1$ is a point.

The local punctual Hilbert scheme $\Hilb_{\prec}^\ell$ at a cusp 
parametrizes ideals of colength $\ell$ in
$\CC[\![\mathtt{t}^2,\mathtt{t}^3]\!](\cong \CC[\![\mathtt{x},\mathtt{y}]\!]/(\mathtt{y}^2-\mathtt{x}^3))$.
If $\ell>1$, such ideals are given by \cite{Pfister:1992dl,Lax:2000xd}
\begin{equation}
\begin{split}
I^\ell(u)&=(\mathtt{t}^\ell+u \mathtt{t}^{\ell+1}),\quad (u\in \CC),\\
Q^\ell&= (\mathtt{t}^{\ell+1},\mathtt{t}^{\ell+2}) 
\end{split}
\end{equation}
with the relation  $\lim_{u\to \infty}I^\ell(u)=Q^\ell$. Hence 
$\Hilb_{\prec}^\ell\cong \PP^1$ if  $\ell>1$.
The only colength one  ideal is $Q^1=(\mathtt{t}^2,\mathtt{t}^3)$.
Thus  $\Hilb_{\prec}^1$ is a point.

Recall that a partition is a sequence $\lambda=(\lambda_1,\lambda_2,\dots)$ of non-negative integers in non-increasing order and 
containing only finitely many non-zero terms. 
We say that  $\lambda$ is a partition of $d$  if $\abs{\lambda}:=\sum\lambda_i=d$.
When $\lambda$ is a partition of $d$, we  use an alternative notation $\lambda=(1^{\delta_1}2^{\delta_2}\cdots d^{\delta_d})$
 where $\delta_\ell=\#\{i\mid \lambda_i=\ell\}$ so that $\sum_{\ell=1}^d  \ell\delta_\ell=d$.

Let $\mathcal{A}$ be the set of nodes on $C$ and $\mathcal{B}$   that of cusps on $C$.  
The argument in \cite{Kawai:2004fh} can be readily extended in the present case and we obtain 
\begin{equation}
\begin{split}
\chi(C^{[d]})=\sum_{\substack{\lambda=(1^{\delta_1}\cdots d^{\delta_d})\\
 \abs{\lambda}=d}}&\sum_{\substack{\mathcal{A}=\sqcup_{\ell=1}^d\mathcal{A}_\ell\\
 \mathcal{B}=\sqcup_{\ell=1}^d\mathcal{B}_\ell\\
 \#\mathcal{A}=a,\, \#\mathcal{B}=b\\
\#\mathcal{A}_\ell +\#\mathcal{B}_\ell=\delta_\ell\  (\ell\ge 2)}}
  \chi\left(\left(C\setminus\sqcup_{\ell=2}^d( \mathcal{A}_\ell\sqcup \mathcal{B}_\ell)\right)^{(\delta_1)}\right)\\
  &\times
  \prod_{\ell=2}^d\left[
  \left(\chi(\Hilb_{ \propto}^\ell)-1\right)^{\#\mathcal{A}_\ell}\left(\chi(\Hilb_{\prec}^\ell)-1\right)^{\#\mathcal{B}_\ell}\right].
\end{split}
\end{equation}
The explicit  descriptions  of  $\Hilb_{ \propto}^\ell$ and $\Hilb_{\prec}^\ell$ in the above
imply that $\chi(\Hilb_{ \propto}^\ell)=2(\ell-1)-(\ell-2)=\ell$  and 
$\chi(\Hilb_{\prec}^\ell)=2$ for $\ell>1$.  Moreover,
\begin{equation}
\chi\left(\left(C\setminus\sqcup_{\ell=2}^d( \mathcal{A}_\ell\sqcup \mathcal{B}_\ell)\right)^{(\delta_1)}\right)
=\binom{\delta_1-1+\chi\left(C\setminus\sqcup_{\ell=2}^d( \mathcal{A}_\ell\sqcup \mathcal{B}_\ell)\right)}{\delta_1}.
\end{equation}
Hence, by setting $a_\ell=\#\mathcal{A}_\ell$ and $b_\ell=\#\mathcal{B}_\ell$, we see that
\begin{equation}
\begin{split}
\chi(C^{[d]})=&\sum_{\substack{\lambda=(1^{\delta_1}\cdots d^{\delta_d}),\,
 \abs{\lambda}=d,\\[2pt]
\sum_{\ell\ge 2}a_\ell  \le a,\\[2pt]  \sum_{\ell\ge 2}b_\ell\le b,\\[2pt]
a_\ell +b_\ell=\delta_\ell\  (\ell\ge 2)}} 
\binom{a}{a-\sum_{\ell\ge 2}a_\ell,a_2,\dots,a_d}\binom{b}{b-\sum_{\ell\ge 2}b_\ell,b_2,\dots,b_d}\\
&\qquad\qquad\times 
\binom{\delta_1-1+\chi(C)-\sum_{\ell\ge 2}(a_\ell+b_\ell)}{\delta_1}
\prod_{\ell=2}^d(\ell-1)^{a_\ell}.
\end{split}
\end{equation}
Now let us switch from  the sum over partitions $\lambda$  to that over $a_\ell$'s and  $b_\ell$'s. Then,
\begin{equation}
  \begin{split}
    &  \chi(C^{[d]})=
\sum_{\substack{a_2\ge 0,\dots,a_d\ge 0\\[2pt]
b_2\ge 0,\dots,b_d\ge 0\\[2pt]
   \sum_{\ell \ge 2} \ell (a_\ell+b_\ell) \le d\\ 
   \sum_{\ell \ge 2} a_\ell\le a\\
    \sum_{\ell \ge 2} b_\ell\le b
   }}
\binom{a}{\sum_{\ell \ge 2} a_\ell}\binom{\sum_{\ell \ge 2} a_\ell}{a_2,\dots,a_d}
\binom{b}{\sum_{\ell \ge 2} b_\ell}\binom{\sum_{\ell \ge 2} b_\ell}{b_2,\dots,b_d}\\
&\qquad\qquad\qquad \times 
\binom{d-\sum_{\ell \ge 2} (\ell+1) (a_\ell+b_\ell)-1+\chi(C)}{d-\sum_{\ell \ge 2} \ell (a_\ell+b_\ell) }          \prod_{\ell=2}^d(\ell-1)^{a_\ell}  \,.
\end{split}
\end{equation}
 Consequently, the generating function becomes
\begin{equation}
 \begin{split}
    & \sum_{d=0}^\infty\chi(C^{[d]})y^d\\
    &\ =\sum_{j=0}^a\,\,\sum_{\substack{a_2\ge 0,a_3\ge 0,\dots\\[2pt]
 j= \sum_{\ell \ge 2} a_\ell}}
\sum_{k=0}^b\,\,\sum_{\substack{b_2\ge 0,b_3\ge 0,\dots\\[2pt]
 k= \sum_{\ell \ge 2} b_\ell}}
  \binom{a}{j}\binom{j}{a_2,a_3,\dots}\binom{b}{k}\binom{k}{b_2,b_3,\dots} \\
&\qquad\qquad\times \left(\prod_{\ell\ge 2}(\ell-1)^{a_\ell}\right)
y^{\sum_{\ell \ge 2} \ell (a_\ell+b_\ell)}(1-y)^{\sum_{\ell\ge 2}(a_\ell+b_\ell)-\chi(C)}\\
&\ =(1-y)^{-\chi(C)}\sum_{j=0}^a\binom{a}{j}\sum_{\substack{a_2\ge 0,a_3\ge 0,\dots\\[2pt]
 j= \sum_{\ell \ge 2} a_\ell}}\binom{j}{a_2,a_3,\dots}
 \prod_{\ell\ge 2}\left\{   (\ell-1)y^\ell(1-y)\right\}^{a_\ell}\\
&\qquad\qquad\qquad\times   \sum_{k=0}^b\binom{b}{k}\sum_{\substack{b_2\ge 0,b_3\ge 0,\dots\\[2pt]
 k= \sum_{m \ge 2} b_m}}\binom{k}{b_2,b_3,\dots}
 \prod_{m\ge 2}\left\{   y^{m}(1-y)\right\}^{b_{m}}
\end{split}
\end{equation}
where we have used the binomial theorem in the first step.
The multinomial theorem further simplifies the last expression  as
\begin{equation}
(1-y)^{-\chi(C)}\sum_{j=0}^a\binom{a}{j}\left(\sum_{\ell \ge 2}(\ell-1)y^\ell(1-y)\right)^j
\sum_{k=0}^b\binom{b}{k}\left(\sum_{m \ge 2}y^m(1-y)\right)^k.
\end{equation}
Hence, by summing  over $\ell$ and $m$ we obtain that
\begin{equation}
\sum_{d=0}^\infty\chi(C^{[d]})y^d=(1-y)^{-\chi(C)}\sum_{j=0}^a\binom{a}{j}\left(  \frac{y^2}{1-y}  \right)^j
\sum_{k=0}^b\binom{b}{k}y^{2k}.
\end{equation}
Finally,  the sums over $j$ and $k$ can be done by the binomial theorem:
\begin{equation}
 \sum_{d=0}^\infty\chi(C^{[d]})y^d
=(1-y)^{-\chi(C)}\left(1+\frac{y^2}{1-y}\right)^a(1+y^2)^b.
\end{equation}
By using 
$\chi(C)=2-2g+a+2b$  one immediately recognizes that this is equivalent to  \eqref{formula}.

%%%%%
\section{Reconciliation with the Gopakumar-Vafa picture}
%%%%%

Let $\nu:\tilde C\to C$ be the normalization.
The generalized Jacobian $J(C)$ fits into an exact sequence of abelian algebraic groups
\begin{equation}\label{genJac}
1\to (\GG_m)^a\times(\GG_a)^b \to J(C) \overset{\nu^*}{\to} J(\tilde C) \to 1 
\end{equation}
where $\GG_m\cong \CC^\times$ is the multiplicative group, $\GG_a\cong \CC$ is the additive group,  $1$ is the trivial group, and $J(\tilde C)$
 is the Jacobian of $\tilde C$. 
Thus to obtain the compactified Jacobian  $\bar J(C)$ from  $J(C)$ one needs  appropriate compactifications of  $\GG_m$ and $\GG_a$.
Let $R_{\propto}$ be a rational curve with a  node,  $R_{\prec}$ a rational curve with a cusp.  We know
 that the nonsingular parts of  $ R_{\propto}$   and $ R_{\prec}$  are respectively isomorphic to $\GG_m$  and $\GG_a$ \cite{Silverman:1992ay}.  
 Hence $ R_{\propto}$ and $ R_{\prec}$  can be regarded as  such compactifications.

To compare our result with the proposal by  Gopakumar and Vafa \cite{Gopakumar:1998qr}  we need to know the ``Lefschetz  $\mathfrak{sl}_2$ action" on $H^*(\bar J(C))$.   At this stage one might worry about the feasibility of this since the so-called ``K\" ahler package"  does not necessarily hold for the usual cohomologies of singular varieties. However, in the present case we may evade this obstacle  by using  the following argument.
 The  curve $R_{\propto}$  is   obtained by shrinking  one of  the two generators of $H_1(E)$ of  an elliptic curve $E$.  Similarly, $R_{\prec}$
 is  obtained by shrinking  both of the two  generators of $H_1(E)$. 
 So, although  $R_{\propto}$ and  $R_{\prec}$ are singular,    $H^*(R_{\propto} )$  and  $H^*( R_{\prec})$ may  still be regarded as the  $\mathfrak{sl}_2$ modules obtained by deleting respectively one spin $0$ and two spin $0$  representations from the  $\mathfrak{sl}_2$ module $H^*(E)$.
 With this interpretation in mind, we have
 \begin{equation}\label{ag1}
\begin{split}
\Tr_{H^*( R_{\propto})}(-1)^{\mathsf{H}}y^{\mathsf{H}}&=-(y^{\inv{2}}-y^{-\inv{2}})^{2}-1,\\
\Tr_{H^*( R_{\prec})}(-1)^{\mathsf{H}}y^{\mathsf{H}}&=-(y^{\inv{2}}-y^{-\inv{2}})^{2}-2.
\end{split}
\end{equation}
(Recall that the arithmetic genera of  $R_{\propto}$, $R_{\prec}$ and $E$ are all equal to one.)
Since the genus of $\tilde C$ is $g-a-b$, it follows from \eqref{genJac}  and \eqref{ag1} that
\begin{equation}
\begin{split}
&\Tr_{H^*(\bar J(C))}(-1)^{\mathsf{H}}y^{\mathsf{H}}\\
&\quad =(-1)^g(y^{\inv{2}}-y^{-\inv{2}})^{2(g-a-b)}\left\{(y^{\inv{2}}-y^{-\inv{2}})^{2}+1\right\}^a
\left\{(y^{\inv{2}}-y^{-\inv{2}})^{2}+2\right\}^b.
\end{split}
\end{equation}
Hence we conclude that  the expected relation indeed holds:
\begin{equation}
(-1)^g\sum_{d=0}^\infty \chi(C^{[d]})y^{d+1-g}=\frac{\Tr_{H^*(\bar J(C))}(-1)^{\mathsf{H}}y^{\mathsf{H}}}{(y^{\inv{2}}-y^{-\inv{2}})^{2}}\,.
\end{equation}
.

\acknowledgments
The author is  supported by KAKENHI (19540024).

\providecommand{\href}[2]{#2}\begingroup\raggedright\endgroup

\end{document}